\documentclass[sigconf,main]{acmart}

\usepackage{multirow}
\usepackage{microtype}

\AtBeginDocument{%
  \providecommand\BibTeX{{%
    \normalfont B\kern-0.5em{\scshape i\kern-0.25em b}\kern-0.8em\TeX}}}

\setcopyright{acmcopyright}
\copyrightyear{2023}
\acmYear{2023}
\acmDOI{XXXXXXX.XXXXXXX}

\acmConference[Conference acronym 'XX]{Make sure to enter the correct
  conference title from your rights confirmation email}{June 03--05,
  2018}{Woodstock, NY}
\acmPrice{15.00}
\acmISBN{978-1-4503-XXXX-X/18/06}

\begin{document}

\title{TweetInfo: An Interactive System to Mitigate Online Harm
}


\author{Gautam Kishore Shahi}
\affiliation{%
  \institution{University of Duisburg-Essen, Duisburg}
  \country{Germany}
    }
    \email{gautam.shahi@uni-due.de}





\renewcommand{\shortauthors}{Anonymous Author(s)}

\begin{abstract}
The increase in active users on social networking sites (SNSs) has also observed an increase in harmful content on social media sites. Harmful content is described as an inappropriate activity to harm or deceive an individual or a group of users. Alongside existing methods to detect misinformation and hate speech, users still need to be well-informed about the harmfulness of the content on SNSs. This study proposes a user-interactive system TweetInfo for mitigating the consumption of harmful content by providing metainformation about the posts. It focuses on two types of harmful content: hate speech and misinformation. TweetInfo provides insights into tweets by doing content analysis. Based on previous research, we have selected a list of metainformation. We offer the option to filter content based on metainformation \textit{Bot, Hate Speech, Misinformation, Verified Account, Sentiment, Tweet Category, Language}. The proposed user interface allows customising the user's timeline to mitigate harmful content. This study present the demo version of the propose user interface of TweetInfo. 

%

\end{abstract}



\maketitle

\vspace{-2mm}
\section{Introduction}

Active social media users have grown significantly over the last decade. Social media has become a source of information for most people,  especially Twitter and Facebook \citep{10.1145/3341161.3343523}, which contains both harmful and non-harmful content. Harmful content spreads faster than non-harmful content, and these posts have a wider readability audience \citep{10.1145/3292522.3326034}. For users, it is difficult to figure out the truthfulness of the content \citep{atodiresei2018identifying}. 

Social media allows users to create user-generated content (UGC) without having strict guidelines \citep{kaplan2010users}. Users browse the content and share it without checking the authenticity of the content \citep{luca2015user}. Social media platforms are well known for spreading harmful content like misinformation or hate speech \citep{shahi2021exploratory,rochert2021networked,kazemi2022research,shahi2020fakecovid, nandini2023explaining}. SNSs provide worldwide access to multimodal user-generated content \citep{shahi2022amused}.
For users, it is hard to verify the authenticity of the information \citep{atodiresei2018identifying}. With the current social media interface, users do not have the option to customise the content for browsing. Hence, users often interact with harmful content \citep{shahi2023exploratory} and spread it through shares, likes, and comments. Sometimes, it causes social crimes like mob lynching and cyberbullying \citep{keipi2016online}. SNSs such as Twitter, Facebook, and YouTube are working on strategies to minimise harmful content on their platforms, respecting free speech for all \citep{8615517}. 


Previous research for hate speech detection and misinformation classifications is limited to a sample dataset, platform, and language. The research has not been transferred to flag the content on social media sites. Thus users need to be made aware of the truthfulness of the UGC. To solve this problem, we proposed an interactive system called TweetInfo for harmful flag content so that users can filter the content on social media sites. It will create awareness and reduce the consumption of harmful content.
 
\begin{figure}
    \centering
    \includegraphics[width=.46\textwidth]{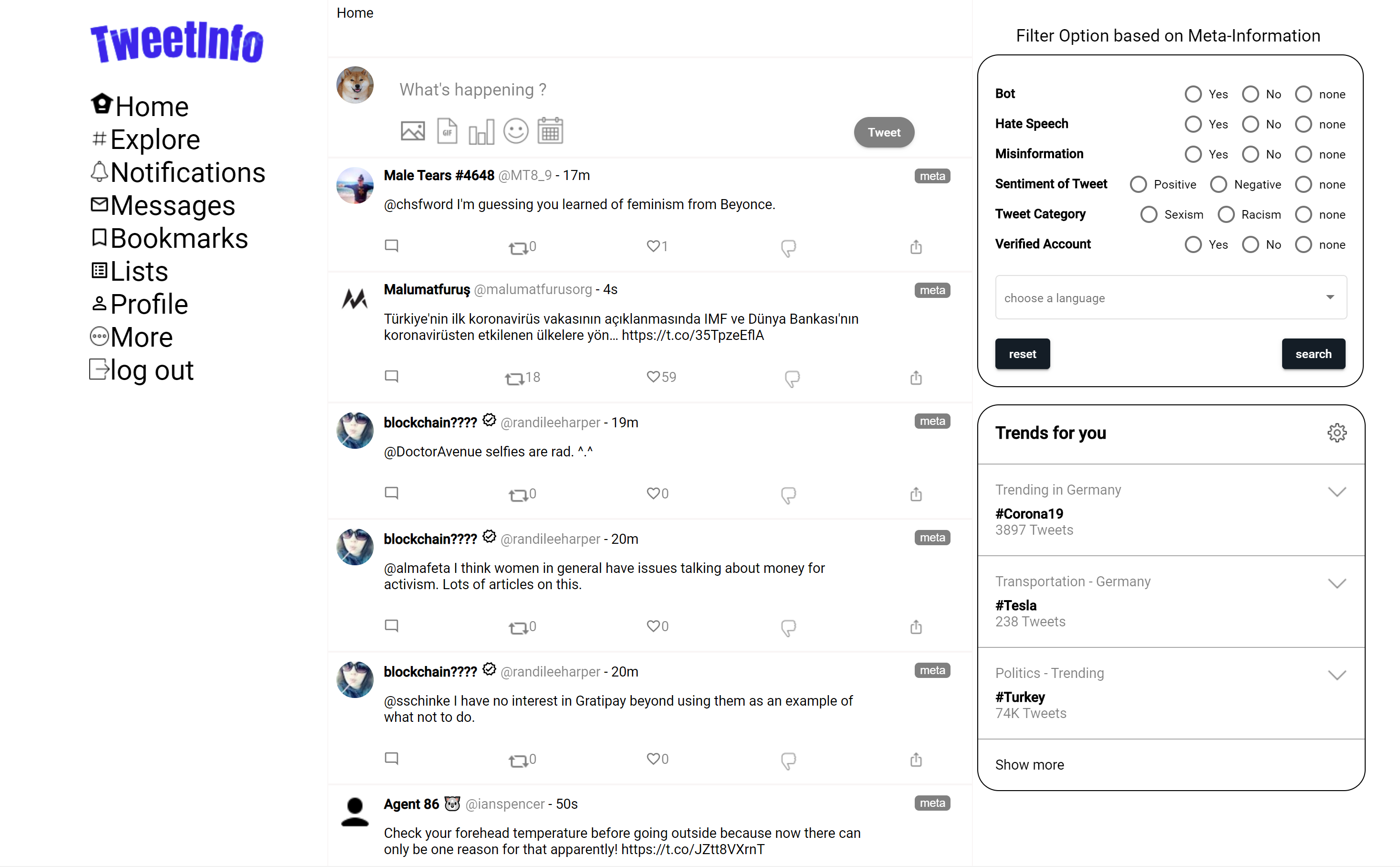}
    \caption{TweetInfo: an interactive system
for visualisation of tweets}
    \label{figure:tweetinfo}
\end{figure}

We first analyse the element that needs to be indicated to build a TweetInfo social media platform. TweetInfo allows users to filter the content as per their choice and interact with the social media post called TweetInfo, which filters the posts and provides meta-information about the social media post based on content analysis. 
A glimpse of TweetInfo is shown in figure~\ref{figure:tweetinfo}.

\vspace{0.00mm}

\vspace{-2mm}
\section{TweetInfo}

For TweetInfo, the front-end contains the meta-information in the pop window and a customisation window. As a back-end, there will be a database containing the labelled tweets.


\textbf{Data Collection} For this study, we have identified the hate speech and misinformation datasets from two different datasets. For hate speech, we have used data from the abusive dataset \citep{waseem-hovy-2016-hateful}, which provides manually annotated data for hate speech on racism and sexism. We have used data from COVID-19 misinformation tweets for misinformation \citep{shahi2021exploratory}. From both datasets, we filtered 989 tweets to use in the study.

\vspace{-3mm}
\subsection{Frontend}

This section will briefly describe how the designed web application operates. Divided into two sections using Flask (Python) for back-end development and Angular 8 for browser view Graphical User Interface. The choice of Flask and Angular for implementing this web application was based on personal experience and the fact that their code maintenance is straightforward due to their extensive use in producing web applications.
To display all attributes obtained above, the MatDialog service from Angular will help to design meta-information. The meta-information will be displayed as a pop-up window after clicking on the" meta" button next to the header of each tweet.

\vspace{-3mm}
\subsection{Customizations}

In this step, we prepare the list of meta-information. We extract the information from the collected data. Based on the prior research, we have used \textit{hate speech, misinformation, bot, languages, gender, sentiment, and verified source} as the set of meta-information. We used the label from the collected dataset to classify hate speech and misinformation. We have calculated the sentiment for both datasets. For sentiment analysis, we have used Vadersentiment analysis\footnote{https://pypi.org/project/vaderSentiment/}.
The score is computed by summing each word's valence scores in the lexicon, adjusted according to the rules, and then normalised to -1 (most extreme negative) and +1 (most extreme positive). Nowadays, social media platforms are multilingual. Currently, language allows filtering tweets in two languages only, i.e., English and Spanish. Social Bot is an agent that communicates more or less likely an original account. Bots are programmed for both good and harmful activity. We identified the bots using BotoMeter API\footnote{https://botometer.osome.iu.edu/api}.
For the Verified Account, Twitter has an account verification feature. Usually, Twitter checks the popular account and verifies it based on the given information. If the profile is verified, it shows a blue tick mark; We used it as metainformation. If the given tweets are misinformation, then we also link to the fact-checked article to provide the background information about the claim in the tweets.

\begin{figure}
    \centering
    \includegraphics[width=.46\textwidth]{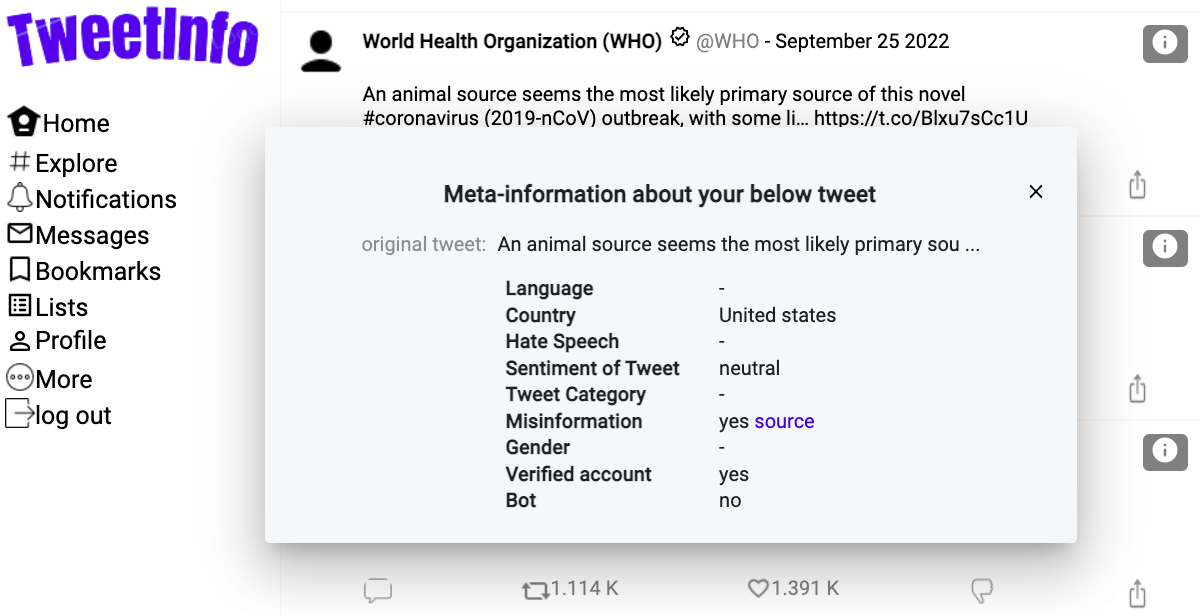}
    \caption{An example of Metainformation generated for a Tweet}
    \label{figure:meta}
\end{figure}

Based on the above metainformation, TweetInfo will enable the selection
of meta-information proposed to filter content on their timeline. After filtering the content, a user can click on the icon to get the metainformation, as shown in figure~\ref{figure:meta}.

\vspace{0.00mm}

\vspace{-3mm}
\subsection{Backend}

We have stored the dataset in the PostgreSQL database; we used Python Flask to connect the database to the front end. We have hosted the web application using Heroku\footnote{www.heroku.com}. With these tools, TweetInfo is running live for the demonstration. TweetInfo is available for the community to test and share their idea. The TweetInfo is access controlled; once a user signs in, we record the mouse click and store it in the database. The mouse click recording will be used in future analysis.

\vspace{-3mm}
\section{Demonstration}

To filter the tweets, please check the box parameter and click on search. By default, all filter options are set as no. Until now, hate, speech and misinformation are mutually exclusive (So users can not choose yes for both options simultaneously). The list of options for customisation and their description on the user instructions. After selecting the customised option, users can browse the tweets as per their choice. To get an overview of the metainformation; please click on the icon. A detailed description and demo video are available on GitHub\footnote{https://gautamshahi.github.io/tweetinfo/}. 

\vspace{-2mm}
\section{Conclusion \& Future Work}
This paper proposes an interactive system for flagging harmful content on social media. TweetInfo is implemented using the pre-classified data stored in the database. A list of metainformation is provided for flagging harmful content on the timeline. In future work, we will conduct a user study to measure the effectiveness of the proposed platform. The idea of TweetInfo can be transferred using a browser plugin to flag the content in real-time on different social media sites. 

\vspace{-3mm}
\bibliographystyle{ACM-Reference-Format}
\bibliography{unreliable-news}


\begin{thebibliography}{15}


\ifx \showCODEN    \undefined \def \showCODEN     #1{\unskip}     \fi
\ifx \showDOI      \undefined \def \showDOI       #1{#1}\fi
\ifx \showISBNx    \undefined \def \showISBNx     #1{\unskip}     \fi
\ifx \showISBNxiii \undefined \def \showISBNxiii  #1{\unskip}     \fi
\ifx \showISSN     \undefined \def \showISSN      #1{\unskip}     \fi
\ifx \showLCCN     \undefined \def \showLCCN      #1{\unskip}     \fi
\ifx \shownote     \undefined \def \shownote      #1{#1}          \fi
\ifx \showarticletitle \undefined \def \showarticletitle #1{#1}   \fi
\ifx \showURL      \undefined \def \showURL       {\relax}        \fi
\providecommand\bibfield[2]{#2}
\providecommand\bibinfo[2]{#2}
\providecommand\natexlab[1]{#1}
\providecommand\showeprint[2][]{arXiv:#2}

\bibitem[Atodiresei et~al\mbox{.}(2018)]%
        {atodiresei2018identifying}
\bibfield{author}{\bibinfo{person}{Costel-Sergiu Atodiresei}, \bibinfo{person}{Alexandru T{\u{a}}n{\u{a}}selea}, {and} \bibinfo{person}{Adrian Iftene}.} \bibinfo{year}{2018}\natexlab{}.
\newblock \showarticletitle{Identifying fake news and fake users on twitter}.
\newblock \bibinfo{journal}{\emph{Procedia Computer Science}}  \bibinfo{volume}{126} (\bibinfo{year}{2018}), \bibinfo{pages}{451--461}.
\newblock


\bibitem[Kaplan and Haenlein(2010)]%
        {kaplan2010users}
\bibfield{author}{\bibinfo{person}{Andreas~M Kaplan} {and} \bibinfo{person}{Michael Haenlein}.} \bibinfo{year}{2010}\natexlab{}.
\newblock \showarticletitle{Users of the world, unite! The challenges and opportunities of Social Media}.
\newblock \bibinfo{journal}{\emph{Business horizons}} \bibinfo{volume}{53}, \bibinfo{number}{1} (\bibinfo{year}{2010}), \bibinfo{pages}{59--68}.
\newblock


\bibitem[Kazemi et~al\mbox{.}(2022)]%
        {kazemi2022research}
\bibfield{author}{\bibinfo{person}{Ashkan Kazemi}, \bibinfo{person}{Kiran Garimella}, \bibinfo{person}{Gautam~Kishore Shahi}, \bibinfo{person}{Devin Gaffney}, {and} \bibinfo{person}{Scott~A Hale}.} \bibinfo{year}{2022}\natexlab{}.
\newblock \showarticletitle{Research note: Tiplines to uncover misinformation on encrypted platforms: A case study of the 2019 Indian general election on WhatsApp}.
\newblock \bibinfo{journal}{\emph{Harvard Kennedy School Misinformation Review}} (\bibinfo{year}{2022}).
\newblock


\bibitem[Keipi et~al\mbox{.}(2016)]%
        {keipi2016online}
\bibfield{author}{\bibinfo{person}{Teo Keipi}, \bibinfo{person}{Matti N{\"a}si}, \bibinfo{person}{Atte Oksanen}, {and} \bibinfo{person}{Pekka R{\"a}s{\"a}nen}.} \bibinfo{year}{2016}\natexlab{}.
\newblock \bibinfo{booktitle}{\emph{Online hate and harmful content: Cross-national perspectives}}.
\newblock \bibinfo{publisher}{Taylor \& Francis}.
\newblock


\bibitem[Lappas et~al\mbox{.}(2019)]%
        {10.1145/3341161.3343523}
\bibfield{author}{\bibinfo{person}{Dimitrios Lappas}, \bibinfo{person}{Panagiotis Karampelas}, {and} \bibinfo{person}{George Fessakis}.} \bibinfo{year}{2019}\natexlab{}.
\newblock \showarticletitle{The Role of Social Media Surveillance in Search and Rescue Missions}. In \bibinfo{booktitle}{\emph{Proceedings of the 2019 IEEE/ACM International Conference on Advances in Social Networks Analysis and Mining}} (Vancouver, British Columbia, Canada) \emph{(\bibinfo{series}{ASONAM ’19})}. \bibinfo{publisher}{Association for Computing Machinery}, \bibinfo{address}{New York, NY, USA}, \bibinfo{pages}{1105–1111}.
\newblock
\showISBNx{9781450368681}
\urldef\tempurl%
\url{https://doi.org/10.1145/3341161.3343523}
\showDOI{\tempurl}


\bibitem[Luca(2015)]%
        {luca2015user}
\bibfield{author}{\bibinfo{person}{Michael Luca}.} \bibinfo{year}{2015}\natexlab{}.
\newblock \showarticletitle{User-generated content and social media}.
\newblock In \bibinfo{booktitle}{\emph{Handbook of media Economics}}. Vol.~\bibinfo{volume}{1}. \bibinfo{publisher}{Elsevier}, \bibinfo{pages}{563--592}.
\newblock


\bibitem[Mathew et~al\mbox{.}(2019)]%
        {10.1145/3292522.3326034}
\bibfield{author}{\bibinfo{person}{Binny Mathew}, \bibinfo{person}{Ritam Dutt}, \bibinfo{person}{Pawan Goyal}, {and} \bibinfo{person}{Animesh Mukherjee}.} \bibinfo{year}{2019}\natexlab{}.
\newblock \showarticletitle{Spread of Hate Speech in Online Social Media}. In \bibinfo{booktitle}{\emph{Proceedings of the 10th ACM Conference on Web Science}} (Boston, Massachusetts, USA) \emph{(\bibinfo{series}{WebSci ’19})}. \bibinfo{publisher}{Association for Computing Machinery}, \bibinfo{address}{New York, NY, USA}, \bibinfo{pages}{173–182}.
\newblock
\showISBNx{9781450362023}
\urldef\tempurl%
\url{https://doi.org/10.1145/3292522.3326034}
\showDOI{\tempurl}


\bibitem[Nandini and Schmid(2023)]%
        {nandini2023explaining}
\bibfield{author}{\bibinfo{person}{Durgesh Nandini} {and} \bibinfo{person}{Ute Schmid}.} \bibinfo{year}{2023}\natexlab{}.
\newblock \showarticletitle{Explaining Hate Speech Classification with Model Agnostic Methods}.
\newblock \bibinfo{journal}{\emph{arXiv preprint arXiv:2306.00021}} (\bibinfo{year}{2023}).
\newblock


\bibitem[R{\"o}chert et~al\mbox{.}(2021)]%
        {rochert2021networked}
\bibfield{author}{\bibinfo{person}{Daniel R{\"o}chert}, \bibinfo{person}{Gautam~Kishore Shahi}, \bibinfo{person}{German Neubaum}, \bibinfo{person}{Bj{\"o}rn Ross}, {and} \bibinfo{person}{Stefan Stieglitz}.} \bibinfo{year}{2021}\natexlab{}.
\newblock \showarticletitle{The Networked Context of COVID-19 Misinformation: Informational Homogeneity on YouTube at the Beginning of the Pandemic}.
\newblock \bibinfo{journal}{\emph{Online Social Networks and Media}}  \bibinfo{volume}{26} (\bibinfo{year}{2021}), \bibinfo{pages}{100164}.
\newblock


\bibitem[{Ruwandika} and {Weerasinghe}(2018)]%
        {8615517}
\bibfield{author}{\bibinfo{person}{N.~D.~T. {Ruwandika}} {and} \bibinfo{person}{A.~R. {Weerasinghe}}.} \bibinfo{year}{2018}\natexlab{}.
\newblock \showarticletitle{Identification of Hate Speech in Social Media}. In \bibinfo{booktitle}{\emph{2018 18th International Conference on Advances in ICT for Emerging Regions (ICTer)}}. \bibinfo{pages}{273--278}.
\newblock


\bibitem[Shahi et~al\mbox{.}(2021)]%
        {shahi2021exploratory}
\bibfield{author}{\bibinfo{person}{Gautam~Kishore Shahi}, \bibinfo{person}{Anne Dirkson}, {and} \bibinfo{person}{Tim~A Majchrzak}.} \bibinfo{year}{2021}\natexlab{}.
\newblock \showarticletitle{An exploratory study of covid-19 misinformation on twitter}.
\newblock \bibinfo{journal}{\emph{Online social networks and media}} (\bibinfo{year}{2021}), \bibinfo{pages}{100104}.
\newblock


\bibitem[Shahi and Majchrzak(2022)]%
        {shahi2022amused}
\bibfield{author}{\bibinfo{person}{Gautam~Kishore Shahi} {and} \bibinfo{person}{Tim~A Majchrzak}.} \bibinfo{year}{2022}\natexlab{}.
\newblock \showarticletitle{AMUSED: An Annotation Framework of Multimodal Social Media Data}. In \bibinfo{booktitle}{\emph{International Conference on Intelligent Technologies and Applications}}. Springer, \bibinfo{pages}{287--299}.
\newblock


\bibitem[Shahi and Majchrzak(2023)]%
        {shahi2023exploratory}
\bibfield{author}{\bibinfo{person}{Gautam~Kishore Shahi} {and} \bibinfo{person}{Tim~A Majchrzak}.} \bibinfo{year}{2023}\natexlab{}.
\newblock \showarticletitle{An Exploratory Study and Prevention Measures of Mob Lynchings: A Case Study of India}.
\newblock In \bibinfo{booktitle}{\emph{PLAIS EuroSymposium on Digital Transformation}}. \bibinfo{publisher}{Springer}, \bibinfo{pages}{103--118}.
\newblock


\bibitem[Shahi and Nandini(2020)]%
        {shahi2020fakecovid}
\bibfield{author}{\bibinfo{person}{Gautam~Kishore Shahi} {and} \bibinfo{person}{Durgesh Nandini}.} \bibinfo{year}{2020}\natexlab{}.
\newblock \showarticletitle{FakeCovid--A multilingual cross-domain fact check news dataset for COVID-19}.
\newblock \bibinfo{journal}{\emph{arXiv preprint arXiv:2006.11343}} (\bibinfo{year}{2020}).
\newblock


\bibitem[Waseem and Hovy(2016)]%
        {waseem-hovy-2016-hateful}
\bibfield{author}{\bibinfo{person}{Zeerak Waseem} {and} \bibinfo{person}{Dirk Hovy}.} \bibinfo{year}{2016}\natexlab{}.
\newblock \showarticletitle{Hateful Symbols or Hateful People? Predictive Features for Hate Speech Detection on Twitter}. In \bibinfo{booktitle}{\emph{Proceedings of the {NAACL} Student Research Workshop}}. \bibinfo{publisher}{Association for Computational Linguistics}, \bibinfo{address}{San Diego, California}, \bibinfo{pages}{88--93}.
\newblock
\urldef\tempurl%
\url{https://doi.org/10.18653/v1/N16-2013}
\showDOI{\tempurl}


\end{thebibliography}

\balance
\end{document}